\documentclass[aps,prfluids,onecolumn,superscriptaddress]{revtex4-2}

\usepackage{graphicx}
\usepackage{amsmath}
\usepackage{amssymb}
\usepackage{xcolor}
\usepackage{tikz}
\usepackage{siunitx}
\usepackage{booktabs} 
\usepackage{multirow} 
\usepackage{bigdelim}
\usepackage{amssymb,pifont}

\definecolor{nltw_blue}{rgb}{0,0.6,1}
\definecolor{sw_green}{rgb}{0,0.5647,0}
\definecolor{hj_orange}{rgb}{0.8118,0.3412,0.2745}

\definecolor{color_1}{rgb}{0.9451,0.5569,0.1098}
\definecolor{color_2}{rgb}{0.8824,0.0980,0}
\definecolor{color_3}{rgb}{0,0.0745,0.7490}
\definecolor{color_4}{rgb}{0.7490,0,0.3765}
\definecolor{color_5}{rgb}{0,0.6000,1.0000}
\definecolor{color_6}{rgb}{0.5490,0.7333,0.1490}
\definecolor{color_7}{rgb}{0.0980,0.5020,0}
\definecolor{color_8}{rgb}{0.25,0.25,0.25}

\DeclareRobustCommand\fullblack {\tikz[baseline=-0.6ex ]\draw[thick] (0,0)--(0.4,0);}
\DeclareRobustCommand\dashedblack {\tikz[baseline=-0.6ex ]\draw[thick,dashed] (0,0)--(0.4,0);}

\newcommand{\PG}[1]{\textcolor{black}{#1}}

\usepackage{subfigure}

\begin{document}


\title{From granular collapses to shallow water waves:\\ A predictive model for tsunami generation}


\author{Wladimir Sarlin}
\email[]{wladimir.sarlin1@universite-paris-saclay.fr}
\affiliation{Universit\'e Paris-Saclay, CNRS, Laboratoire FAST, F-91405 Orsay, France}
\author{Cyprien Morize}
\email[]{cyprien.morize@universite-paris-saclay.fr}
\affiliation{Universit\'e Paris-Saclay, CNRS, Laboratoire FAST, F-91405 Orsay, France}
\author{Alban Sauret}
\affiliation{University of California, Santa Barbara, Department of Mechanical Engineering, CA 93106, USA}
\author{Philippe Gondret}
\affiliation{Universit\'e Paris-Saclay, CNRS, Laboratoire FAST, F-91405 Orsay, France}


\date{\today}

\begin{abstract}

In this article, we present a predictive model for the amplitude of impulse waves generated by the collapse of a granular column into a water layer. The model, which combines the spreading dynamics of the grains and the wave hydrodynamics in shallow water, is successfully compared to a large dataset of laboratory experiments, and captures the influence of the initial parameters while giving an accurate prediction. Furthermore, the role played on the wave generation by two key dimensionless numbers, \textit{i.e.}, the global Froude number and the relative volume of the immersed deposit, is rationalized. These results provide a simplified, yet comprehensive, physical description of the generation of tsunami waves engendered by large-scale subaerial landslides, rockfalls, or cliff collapses in a shallow water.
\end{abstract}


\maketitle



\noindent \textbf{{Introduction.}} Tsunamis are among the most destructive natural disasters for human coastal settlements. While events generated by earthquakes have been extensively studied \cite{1996_tadepalli,2006_kanoglu,2006_synolakis,2011_dutykh}, several past or potential occurrences of high amplitude waves arising from large-scale landslides have also been reported in past decades \cite{2001_ward,2009_fritz,2019_grilli,2021_waldmann,2022_rauter}, which constitutes a grand challenge in environmental fluid mechanics \cite{2021_dauxois}. The 1958 Lituya Bay tsunami, featuring the highest recorded wave runup of 524 m \cite{2009_fritz}, is reminiscent of the importance of understanding the physics underlying such events, for reliable hazard assessments. 

A relevant approach is to experimentally model landslides and pyroclastic flows using a granular material \cite{2006_lajeunesse,2009_fritz,2011_roche,2014_viroulet,2015_langlois,2016_lindstrom,2017_mulligan,2019_bullard,2020_bougouin}. Although the finding of constitutive laws for granular media remains challenging, and is still attracting many research activities \cite{2006_jop,2013_bouzid,2014_henann,2020_seongmin,2020_gaume}, canonical experimental configurations have been developed to study geophysical flows, such as the granular collapse experiment \cite{2004_lajeunesse,2004_lube,2005_lajeunesse,2005_lube,2005_balmforth,2005_zenit,2005_staron,2007_staron,2008_lacaze,2009_lacaze,2011_lagree,2021_man,2021c_sarlin}. In this situation, a column of dry grains, suddenly released, falls vertically while spreading horizontally under the effect of gravity. The resulting final deposit height $H_\infty$ and runout distance $\Delta L_\infty=L_\infty-L_0$ both depend on the aspect ratio $a$ of the column, \textit{i.e.}, the ratio of its initial height $H_0$ to width $L_0$, through nontrivial scaling laws \cite{2005_lajeunesse,2005_lube}. The aspect ratio also influences the characteristic timescale of the horizontal spreading \cite{2008_lacaze,2021c_sarlin}. In the case of an immersed granular collapse, similar scaling laws are obtained for the final morphology of the deposit \cite{2012_topin,2018_bougouin,2018_jing}. An interesting feature of the granular collapse experiment was its ability to compare successfully with large-scale geophysical events \cite{2006_lajeunesse}. 

This experimental configuration has also been used recently to generate impulse waves of geophysical interest, by releasing the grains into a water layer of depth $h_0$ \cite{2020a_huang,2020_cabrera,2021a_robbe-saule,2021b_robbe-saule,2021b_sarlin,2022_nguyen}. Several dimensionless parameters have been observed to be important for the wave generation. The most intuitive is the Froude number, that can be defined in two ways: globally as $\mathrm{Fr}_0=\sqrt{H_0/h_0}$ \cite{2020_cabrera,2020a_huang,2022_nguyen}, which compares the typical vertical free-fall velocity $\sqrt{gH_0}$ of the grains to the velocity $\sqrt{gh_0}$  of linear gravity waves in shallow water, or locally as $\mathrm{Fr}_f=v_m/\sqrt{gh_0}$, with $v_m$ the maximum horizontal velocity of the granular front at the water surface \cite{2021a_robbe-saule,2021b_sarlin}. In particular, the wave amplitude was related analytically to $\mathrm{Fr}_f$ through weakly nonlinear scalings \cite{2021b_sarlin}, or with the linear approximation $A_m/h_0 \simeq 1.2~\mathrm{Fr}_f$ \cite{1970_noda,2021a_robbe-saule}. In addition, the wave amplitude $A_m$ was found to be linked to the final immersed volume of grains $\Delta V_\infty$, which can be measured in the field, and allows back-calculations of former events \cite{2021b_robbe-saule}. However, so far a unifying model relating the wave amplitude to the initial geometry of the column and the water depth remains elusive. \PG{Such an attempt was made recently through the study of the collapse of a Newtonian viscous fluid into a water layer \cite{2022_kriaa}. Nevertheless, this configuration is closer to a dam-break problem and does not account for the granular nature of the landslide. As a result,} the prediction of the wave amplitude following a granular collapse currently relies on empirical fits, \PG{without relevant physical model.}

In the present study, a predictive model for the wave amplitude in shallow water conditions is introduced, by coupling the dynamics of the granular collapse highlighted in Ref. \cite{2021c_sarlin} with the wave hydrodynamics elucidated in Ref. \cite{2021b_sarlin}. The influence of the initial parameters is revealed by the model, which therefore captures the role played by the initial Froude number $\mathrm{Fr}_0$ and also explains how the final volume of immersed grains $\Delta V_\infty$ \PG{is related to} the wave generation. The model is successfully compared with an extended dataset of experiments.\\

\begin{figure}[ht!]
	\centering
	\includegraphics[width=0.49\linewidth]{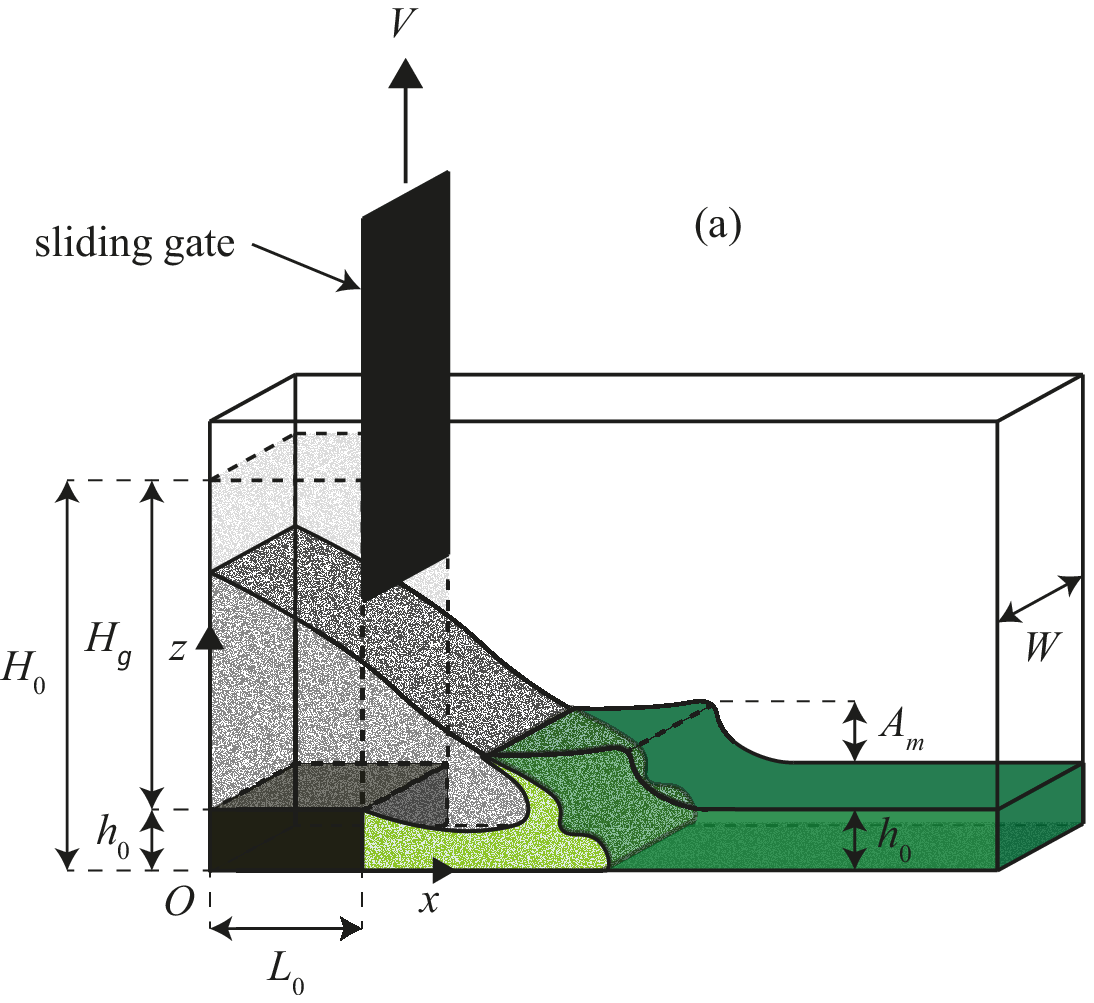}
	\includegraphics[width=0.49\linewidth]{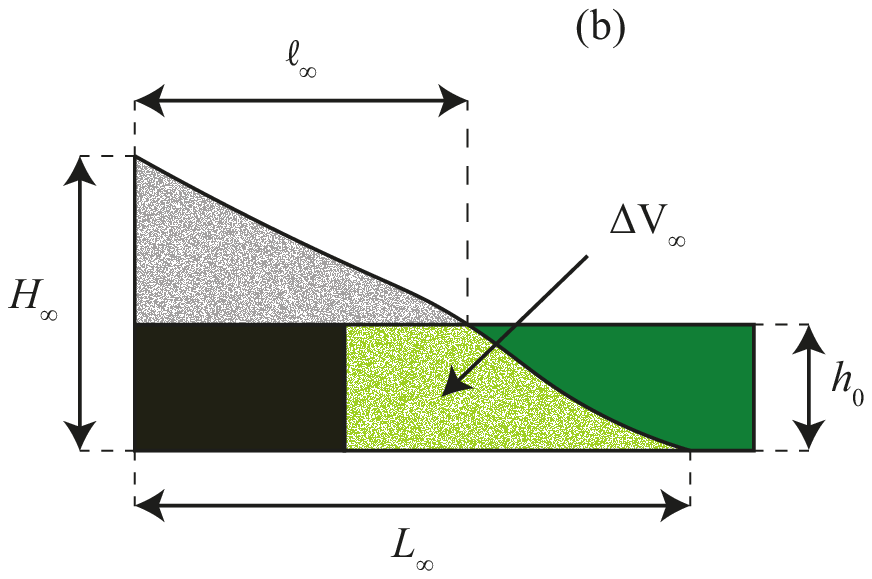}
	\includegraphics[width=0.49\linewidth]{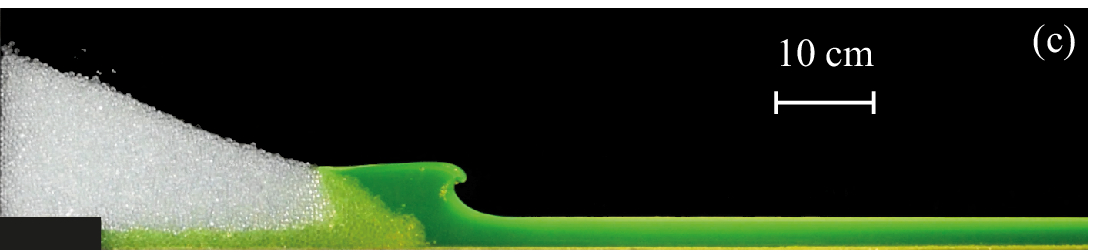}
	\includegraphics[width=0.49\linewidth]{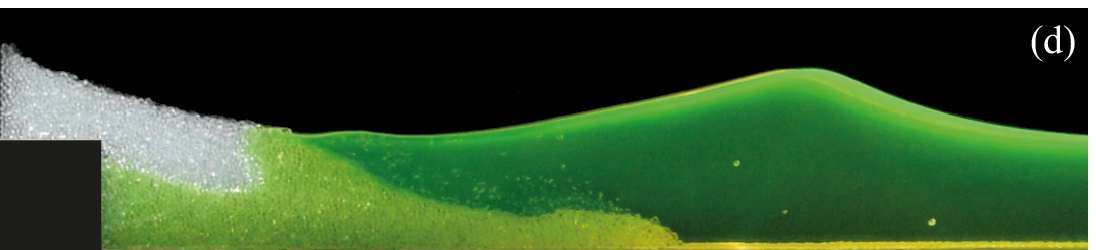}
	\caption{(a) 3D view of the experimental setup, showing the initial granular column of height $H_g$ and width $L_0$, the initial fluid depth $h_0$, the channel width $W$, and the maximum wave amplitude $A_m$. (b) 2D view of the final granular deposit of height $H_\infty$, runout length $L_\infty$, immersed volume $\Delta V_\infty$, and front position $\ell_\infty$ at $z=h_0$. (c)-(d) Impulse waves generated by the collapse of a column with $H_g=39$ cm and $L_0=10$ cm: (c) Bore wave for $h_0=3$ cm, and (d) solitary wave for $h_0=10$ cm (movies are available in Supplemental Material \cite{2022a_sarlin_supplemental_material}).}
	\label{fig:fig1}
\end{figure}

\medskip
\noindent \textbf{{Experimental setup.}} The two-dimensional experimental setup, sketched in figure \ref{fig:fig1}(a), consists in a $2\ \mathrm{m}\times0.30\ \mathrm{m}\times0.15\ \mathrm{m}$ reservoir, filled up to a height $h_0$ with water \cite{2021a_robbe-saule,2021b_sarlin}. A rectangular dry granular column of initial height $H_g$ and width $L_0$ rests on a solid step of height $h_0$, and is retained by a sliding gate. The following ranges of initial parameters were investigated: $9\ \mathrm{cm}\leqslant H_g\leqslant50\ \mathrm{cm}$, $2.5\ \mathrm{cm}\leqslant L_0\leqslant20\ \mathrm{cm}$, and $2\ \mathrm{cm}\leqslant h_0\leqslant25\ \mathrm{cm}$. The total height $H_0=H_g+h_0$ from the bottom plane and the aspect ratio $a=H_0/L_0$ of the initial column were thus varied in the ranges $11\ \mathrm{cm} \leqslant H_0 \leqslant 59\ \mathrm{cm}$ and $1 \lesssim a \lesssim 21$, respectively. This allowed to explore a larger range of initial parameters, especially for the aspect ratio, than in previous studies \cite{2020_cabrera,2021a_robbe-saule,2021b_sarlin}. The grains are monodisperse glass beads of mean diameter $d = 5$ mm, density $\rho = 2.5\ \rm{g.cm^{-3}}$ with a dense packing fraction $\phi \simeq 0.64$. Using the same grains for all experiments is motivated by a previous study \cite{2021a_robbe-saule}, which showed that the maximum amplitude of the generated wave does not significantly depend on the grains' size and density, at least for millimeter-scale grains denser than water. At the beginning of the experiment, the gate is quickly lifted at $V=1\ \rm{m.s^{-1}}$ to ensure no significant influence of the release process on the collapse dynamics \cite{2021c_sarlin}. The solid step restricts fluid perturbations induced by the withdrawal of the gate \cite{2020_cabrera}. The grains fall under the effect of gravity into water, leading to the formation of an impulse wave, as illustrated in figures \ref{fig:fig1}(c)-(d). The video recordings are then processed to extract the maximum amplitude $A_m$ reached by the wave, and the maximum horizontal velocity $v_m=\mathrm{d} \ell / \mathrm{d} t$ of the advancing granular front at the water surface $z=h_0$ \cite{2021b_sarlin}. In addition, the final morphology of the deposit is characterized by measuring the final runout distance $L_\infty$, height $H_\infty$, volume $\Delta V_\infty$ of immersed grains, and position $\ell_\infty$ of the granular front at $z=h_0$, as illustrated in figure \ref{fig:fig1}(b).

\begin{figure}[t]
	\centering									\includegraphics[width=0.75\linewidth]{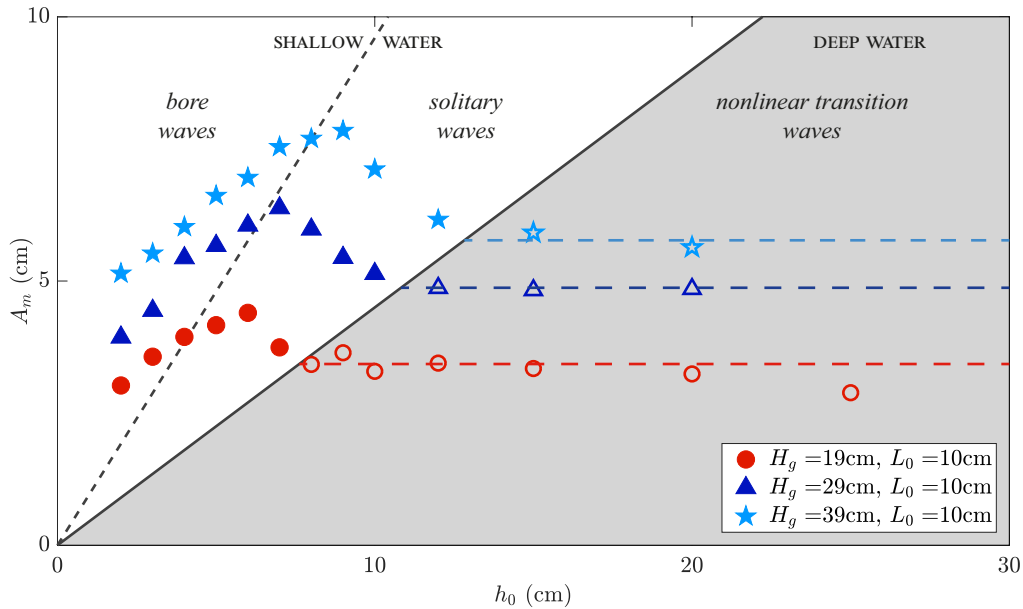}
	
	\caption{Evolution of the maximum wave amplitude $A_m$ with the fluid depth $h_0$, for three different initial granular columns of same width $L_0$ but different height $H_g$. The horizontal dashed lines indicate the saturation of $A_m$ at large $h_0$. Full symbols refer to shallow water waves, while empty symbols correspond to deep water waves. The shaded area corresponds to deep water waves, where $A_m \lesssim 0.45~h_0$, and the black dashed line (\dashedblack) represents the \PG{observed} limit between bore and solitary waves ($A_m \simeq 0.96~h_0$).}
	\label{fig:fig2}
\end{figure}

\begin{figure}[t]
	\centering
	\includegraphics[width=0.49\linewidth]{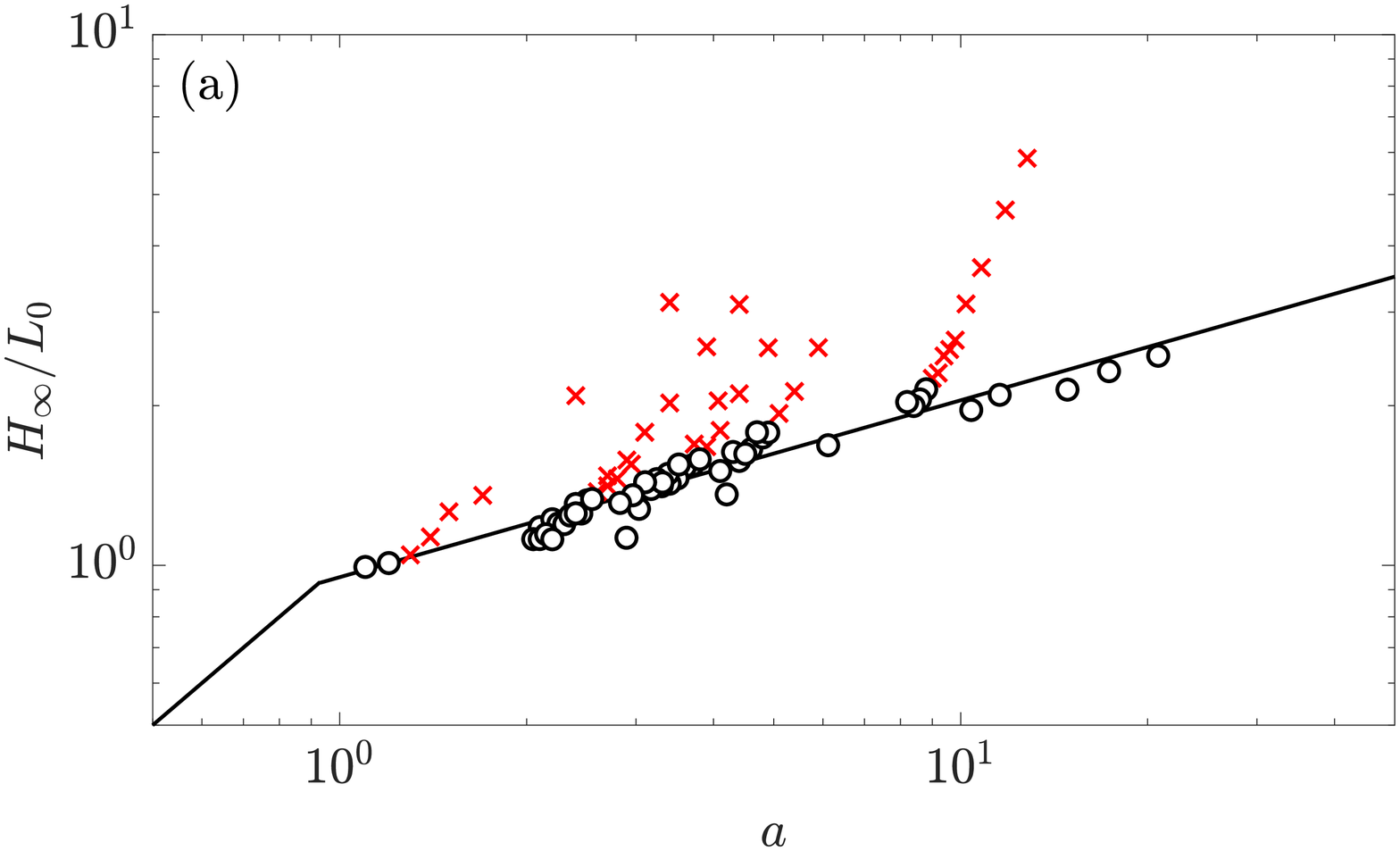}
	\hfill
	\includegraphics[width=0.49\linewidth]{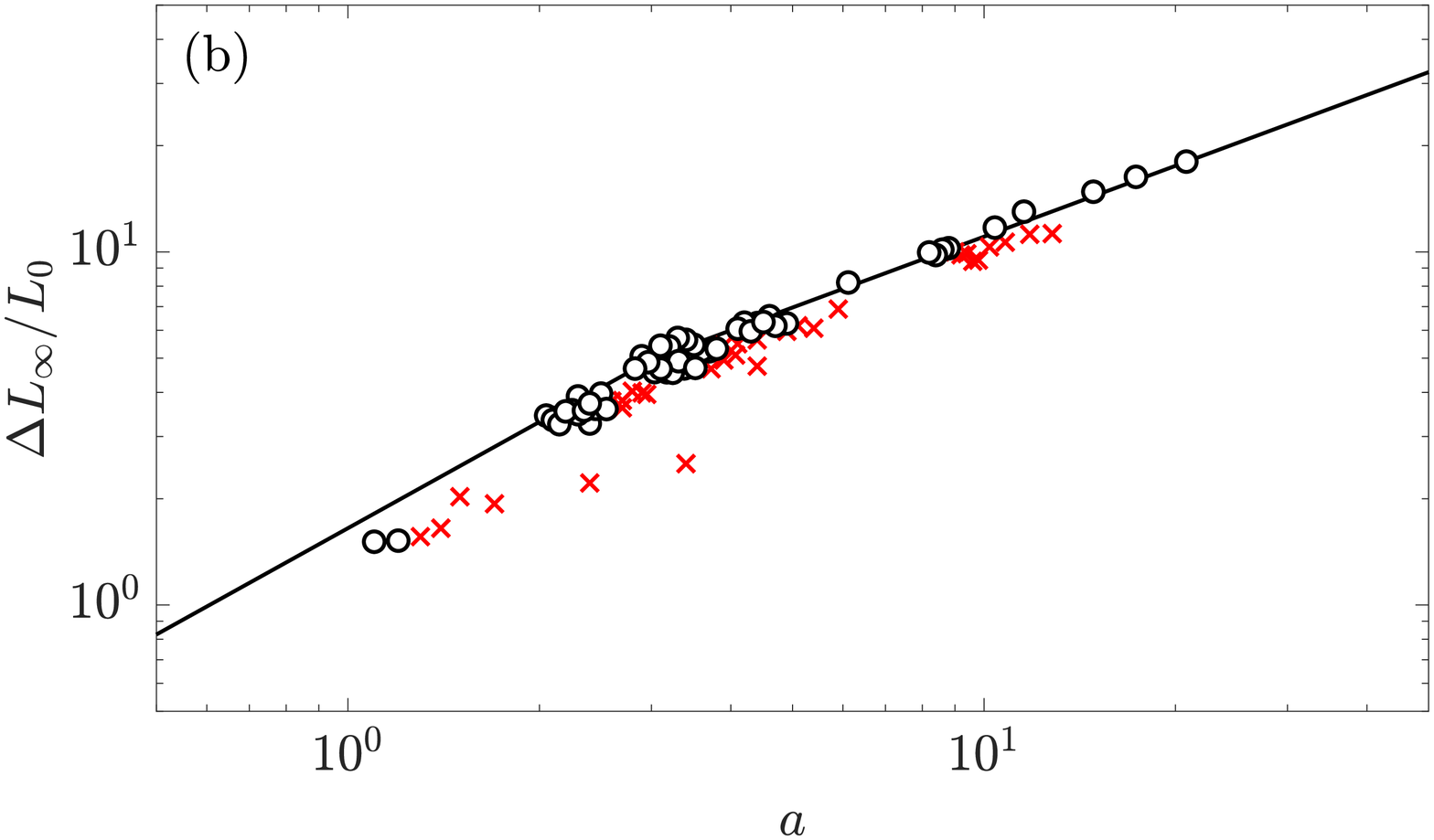}
	\caption{(a) Relative final height $H_\infty/L_0$ and (b) runout distance $\Delta L_\infty/L_0$ reached by the final deposit as a function of the aspect ratio of the initial column $a=H_0/L_0$. (\fullblack) Power laws $H_\infty/L_0=\alpha a^n$ [with $\alpha=1$ and $n=1$ when $a \lesssim 0.93$, and $\alpha=0.95$ and $n=1/3$ when $a \gtrsim 0.93$], and $\Delta L_\infty/L_0=\beta a^m$ [with $\beta=1.65$ and $m=1$ when $a \lesssim 3$, and $\beta=2.38$ and $m=2/3$ when $a \gtrsim 3$]. ($\boldsymbol{\circ}$) Experiments of bore and solitary waves which are not significantly affected by the presence of the solid step. (\textcolor{red}{$\boldsymbol{\times}$}) Bore and solitary waves with a noticeable influence of the solid step, and nonlinear transition waves.}
	\label{fig:fig6}
\end{figure}

\medskip

\noindent \textbf{{Influence of the initial parameters \PG{on the generated wave and final deposit}.}} The evolution of the maximum wave amplitude $A_m$ with the water depth $h_0$ is presented in figure \ref{fig:fig2} for three initial geometries of the column. For a given column geometry, $A_m$ first increases with $h_0$ at the shallowest depths, then decreases for intermediate depths, before saturating at a constant asymptotic value. When the initial fluid depth $h_0$ is small compared to the column height $H_g$, the collective motion of the spreading grains is mainly horizontal, and the advancing front is similar to a piston pushing the water, as illustrated in figures \ref{fig:fig1}(c)-(d) (see also Supplemental Material \cite{2022a_sarlin_supplemental_material}). This situation leads to the formation of either transient bore ($A_m \gtrsim 0.96~h_0$) or solitary ($0.45~h_0 \lesssim A_m \lesssim 0.96~h_0$) waves \cite{2021b_sarlin}. \PG{In that case, the generation mechanism, \textit{i.e.}, the advancing granular front pushing water like a moving piston, is in agreement with recent numerical simulations which reproduced finely a specific geophysical case with a complex topography \cite{2022_rauter}, although the initial conditions are quite different than from our model laboratory configuration.} On the contrary, when $h_0$ is of the order or greater than $H_g$, the grains have essentially a vertical motion, closer to a ``granular spillway" situation, in which a finite volume of grains discharges into deep water (see Supplemental Material \cite{2022a_sarlin_supplemental_material}). This regime departs from shallow water conditions and leads to the formation of \PG{the so-called} nonlinear transition waves for which $A_m \lesssim 0.45~h_0$ \cite{2021b_sarlin}. Increasing the initial height $H_g$ of the column, while keeping the other initial parameters constant, systematically leads to higher values for the wave amplitude. In the following, the study focuses on shallow water waves ($A_m \gtrsim 0.45~h_0$), whose understanding is of significant importance for risk assessments as bore waves are potentially very destructive in enclosed fjords scenarios \cite{2021_waldmann}, while at the same time solitary waves can propagate and distribute part of the collapse energy far from the source.

\PG{In the present configuration, the generated wave is a direct consequence of the granular collapse. Therefore, it is interesting to focus also on the main features of the final granular deposit, characterized by its final height $H_\infty$ and runout distance $\Delta L_\infty = L_\infty - L_0$. In the dry case, these two final parameters, rescaled by the initial width of the column, $H_\infty/L_0$ and $\Delta L_\infty/L_0$, are known to be governed mainly by the initial aspect ratio $a$ of the column, and slightly influenced by the material and frictional properties of the grains and the substrate \cite{2005_lajeunesse,2005_lube,2005_zenit,2007_staron}. Figures \ref{fig:fig6}(a) and \ref{fig:fig6}(b) show the evolution of $H_\infty/L_0$ and $\Delta L_\infty/L_0$, respectively, as a function of $a$. In the present experiments, the initial granular column stands on a solid step of height $h_0$, and one may expect that the final deposit may be different from the one that would occur without a solid step.}
The initial dataset is thus subdivided into two parts: ($\boldsymbol{\circ}$) the experiments for which the solid step has no significant influence on the final deposit, \textit{i.e.}, where $H_0/h_0 \geqslant 4$ and $h_0 \leqslant L_0$ according to \cite{2019_robbe-saule}, and (\textcolor{red}{$\boldsymbol{\times}$}) those impacted by the presence of the step. For the first set ($\boldsymbol{\circ}$), the relative final height $H_\infty/L_0$ [figure \ref{fig:fig6}(a)], and runout distance $\Delta L_\infty/L_0$ [figure \ref{fig:fig6}(b)] are then fitted by piecewise power laws of the initial aspect ratio, following Lajeunesse \textit{et al.} (2005) \cite{2005_lajeunesse}:

\begin{equation}
\label{eq:final_morphology}
\begin{aligned}
	\frac{H_\infty}{L_0} = \alpha a^{n} = & \left\{ \begin{array}{ll}
	\displaystyle a\\[8pt]
	\displaystyle 0.95~a^{1/3}
	\end{array}\right.
    \begin{array}{ll}
	\displaystyle \ \mathrm{for\ }a \lesssim 0.93,\\[8pt]
	\displaystyle \ \mathrm{for\ }a \gtrsim 0.93,
	\end{array}
	\\\\
	\frac{\Delta L_\infty}{L_0} = \beta a^{m} = & \left\{ \begin{array}{ll}
	\displaystyle 1.65~a\\[8pt]
	\displaystyle 2.38~a^{2/3}
	\end{array}\right.
    \begin{array}{ll}
	\displaystyle \ \mathrm{for\ }a \lesssim 3,\\[8pt]
	\displaystyle \ \mathrm{for\ }a \gtrsim 3,
	\end{array}
\end{aligned}
\end{equation}

\noindent where the prefactors $\alpha$ and $\beta$ are slightly dependent on the material \PG{and frictional properties} (except the trivial value $\alpha = 1$ for low enough aspect ratios) \cite{2005_balmforth,2007_staron,2021_man}, \PG{and would also be impacted by the presence of cohesion \cite{2015_langlois,2021_li}}. \PG{Therefore, using different grain or substrate properties would require one to adjust these coefficients accordingly. The exponents of Eqs. \eqref{eq:final_morphology} differ from 2D to 3D configurations \cite{2005_lube,2005_lajeunesse}, and are a signature of the gravity-driven dynamics of the collapse \cite{2005_staron}.}
It should be mentioned that although no experimental data are available here at low aspect ratio ($a \lesssim 0.93$), the scaling $H_\infty/L_0=a$ is represented, as this case corresponds to the trapezoidal final shape of the deposit already described by previous work \cite{2005_lajeunesse,2005_lube,2005_staron,2011_lagree,2021c_sarlin}, where only a fraction of the initial granular column collapses, so that $H_\infty=H_0$. \PG{The second set (\textcolor{red}{$\boldsymbol{\times}$}) of experiments deviates significantly from the scalings of Eqs. \eqref{eq:final_morphology} for the final height as well as for the run-out distance. These data are either impacted by the presence of the solid step or depart from shallow water conditions (deep water waves), and are therefore not considered in the following.}


\begin{figure}[t]
\centering
\includegraphics[width = 0.32\textwidth]{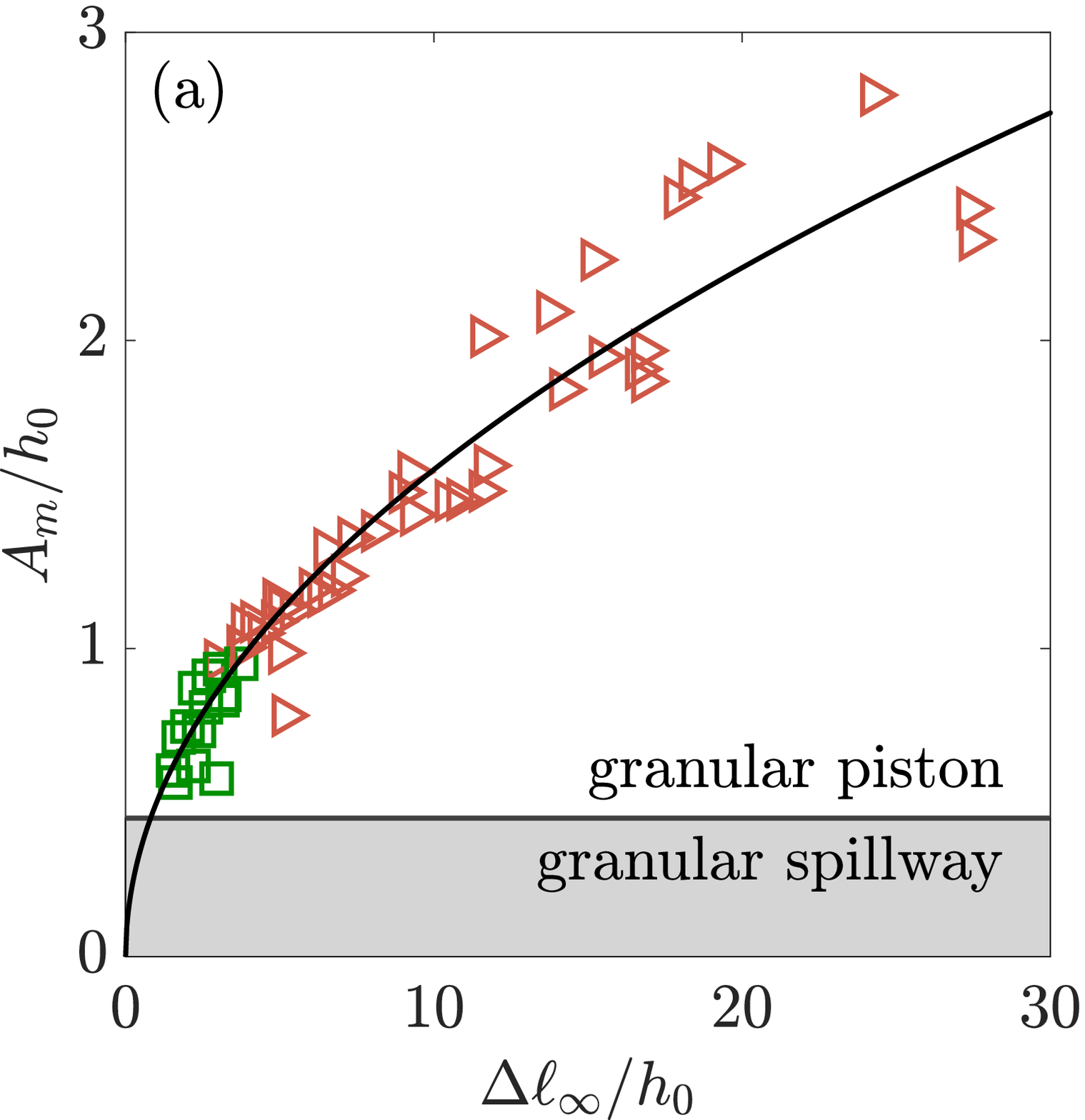}
\hfill
\includegraphics[width = 0.32\textwidth]{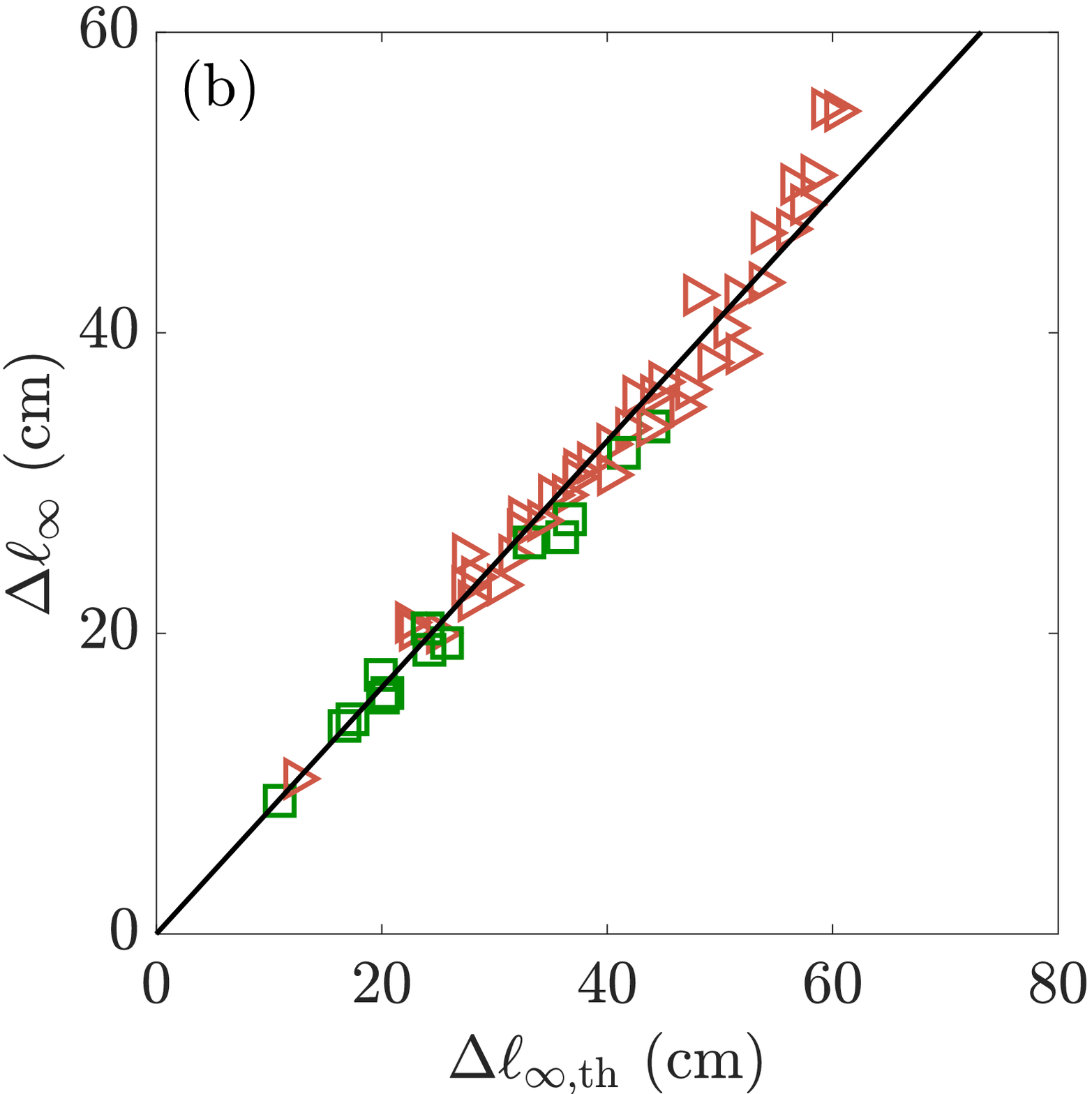}
\hfill
\includegraphics[width = 0.32\textwidth]{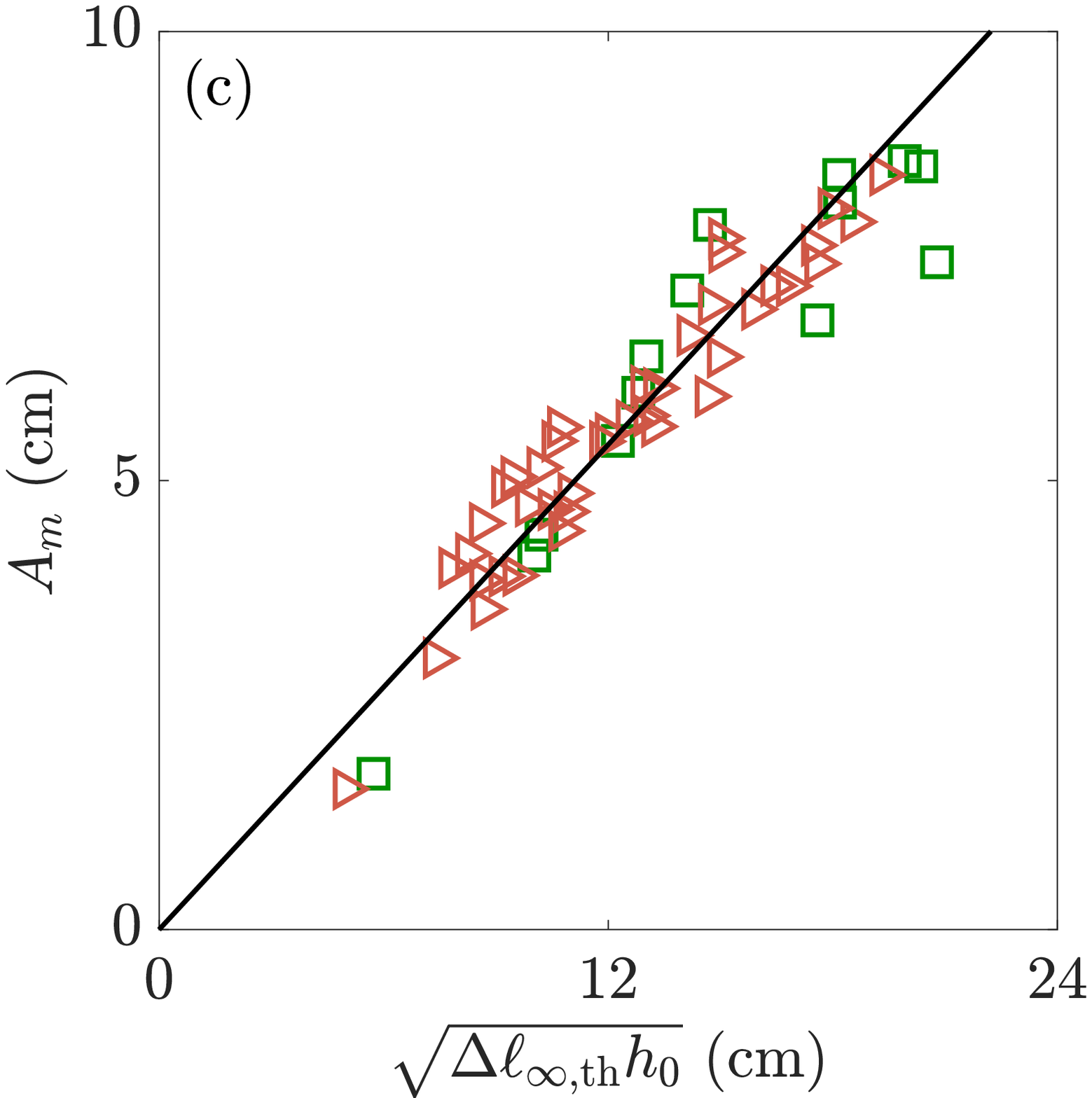}
\caption{(a) Relative maximum amplitude $A_m/h_0$ of the wave as a function of the relative final extension of the granular front $\Delta \ell_{\mathrm{\infty}}/h_0$, with (\fullblack) $A_m/h_0=0.50\sqrt{\Delta \ell_{\mathrm{\infty}}/h_0}$. (b) Comparison between the measurement of $\Delta \ell_{\mathrm{\infty}}$ and the model prediction $\Delta \ell_{\mathrm{\infty,th}}$. (\fullblack) $\Delta \ell_{\mathrm{\infty}}=0.82\Delta \ell_{\mathrm{\infty,th}}$. (c) $A_m$ as a function of $\sqrt{\Delta \ell_{\mathrm{\infty,th}} h_0}$, with  (\fullblack) Eq. \eqref{eq:Am}. Symbols designate (\textcolor{hj_orange}{$\triangleright$}) bore and (\textcolor{sw_green}{$\square$}) solitary waves. The shaded area in (a) correspond to the deep water waves region where $A_m \lesssim 0.45~h_0$.}
\label{fig:fig3}
\end{figure}

\medskip

\noindent \textbf{{A predictive model for the wave amplitude.}} To understand the bell-shaped curves for $A_m(h_0)$ in shallow water, as illustrated in figure \ref{fig:fig2}, the local Froude number $\mathrm{Fr}_f=v_m/\sqrt{gh_0}$ needs to be related to the initial parameters of the granular column as it is known that, at first order, $A_m/h_0 \simeq 1.2~\mathrm{Fr}_f$ \cite{1970_noda,2021a_robbe-saule}. The granular front velocity $v_m$ is expected to scale as $\Delta \ell_\infty/\tau_x$, where $\Delta \ell_\infty=\ell_\infty-L_0$ is the front position of the final deposit and $\tau_x$ the typical spreading time of the granular mass, both taken at $z=h_0$. In a recent study, $\tau_x$ was shown to be proportional to $\sqrt{\Delta \ell_\infty/g}$ \cite{2021c_sarlin}. Using this result, $\mathrm{Fr}_f$ may be estimated as

\begin{equation}
\label{eq:new_local_froude}
\mathrm{Fr}_f = \frac{v_m}{\sqrt{gh_0}} \propto \frac{\Delta \ell_{\infty}}{\tau_x\sqrt{gh_0}} \propto \sqrt{\frac{\Delta \ell_{\infty}}{h_0}},
\end{equation}

\noindent and therefore $A_m/h_0$ is also expected to scale as $\sqrt{\Delta \ell_{\infty}/h_0}$, due to the aforementioned approximate linear relation between $A_m/h_0$ and $\mathrm{Fr}_f$. Figure \ref{fig:fig3}(a) presents the rescaled wave amplitude $A_m/h_0$ as a function of $\Delta \ell_{\infty}/h_0$ for (\textcolor{hj_orange}{$\triangleright$}) bore and (\textcolor{sw_green}{$\square$}) solitary waves. All experimental data \PG{corresponding to shallow water waves} collapse onto a master curve of equation $A_m/h_0 \simeq 0.5 \sqrt{\Delta \ell_{\infty}/h_0}$.
\PG{Therefore, in this case the moving granular front acts like a piston pushing the water, whose velocity would be coupled to its stroke [as highlighted by Eq. \eqref{eq:new_local_froude}].}

An estimate of the typical horizontal extension $\Delta \ell_{\mathrm{\infty,th}}$ can be obtained by assuming a triangular shape for the final deposit: $\Delta \ell_{\mathrm{\infty,th}} = \Delta L_\infty - h_0L_\infty/H_\infty$. Using Eqs. \eqref{eq:final_morphology} the final runout distance at $z=h_0$ is thus expected to be given by the relation

\begin{equation}
	\Delta \ell_{\mathrm{\infty,th}} = \beta a^{m-1} H_0 - \frac{1 + \beta a^m}{\alpha a^n} h_0.
  \label{eq:lth}
\end{equation}

\noindent In figure \ref{fig:fig3}(b), the measured front position $\Delta \ell_\infty$ of the final deposit is compared with the prediction of Eq. \eqref{eq:lth}. The relation $\Delta \ell_\infty=0.82~\Delta \ell_{\mathrm{\infty,th}}$ captures all the experiments, with a prefactor revealing a systematic overestimate of $\Delta \ell_{\infty}$ by the model, as the final deposit is not perfectly triangular, but presents \PG{some curvature} instead \cite{2005_lajeunesse,2005_lube,2005_staron,2021c_sarlin}. From the two fits of figures \ref{fig:fig3}(a) and \ref{fig:fig3}(b), the wave amplitude should be given by

\begin{equation}
\label{eq:Am}
A_m = 0.45\sqrt{\Delta \ell_\mathrm{\infty,th} h_0}.
\end{equation}

\noindent The prediction given by Eq. \eqref{eq:Am} fits well the data, as presented in figure \ref{fig:fig3}(c), which reveals that the model captures the physics behind the wave generation in shallow water. \PG{By inserting Eq. (\ref{eq:lth}) in Eq. (\ref{eq:Am}), the wave amplitude can be expressed as a function of the initial parameters $H_0$, $L_0$, and $h_0$ to obtain}

\begin{equation}
\label{eq:eq:Am_explicit}
A_m = 0.45\,h_0\,\sqrt{\beta \left(\frac{H_0}{L_0}\right)^{m-1} \frac{H_0}{h_0} - \frac{1+\beta (H_0/L_0)^m}{\alpha (H_0/L_0)^n}},
\end{equation}

\noindent where the coefficients $\alpha$, $\beta$, $n$ and $m$ are given in Eqs. (\ref{eq:final_morphology}).

\begin{figure}[t]
	\centering
	\includegraphics[width=0.49\linewidth]{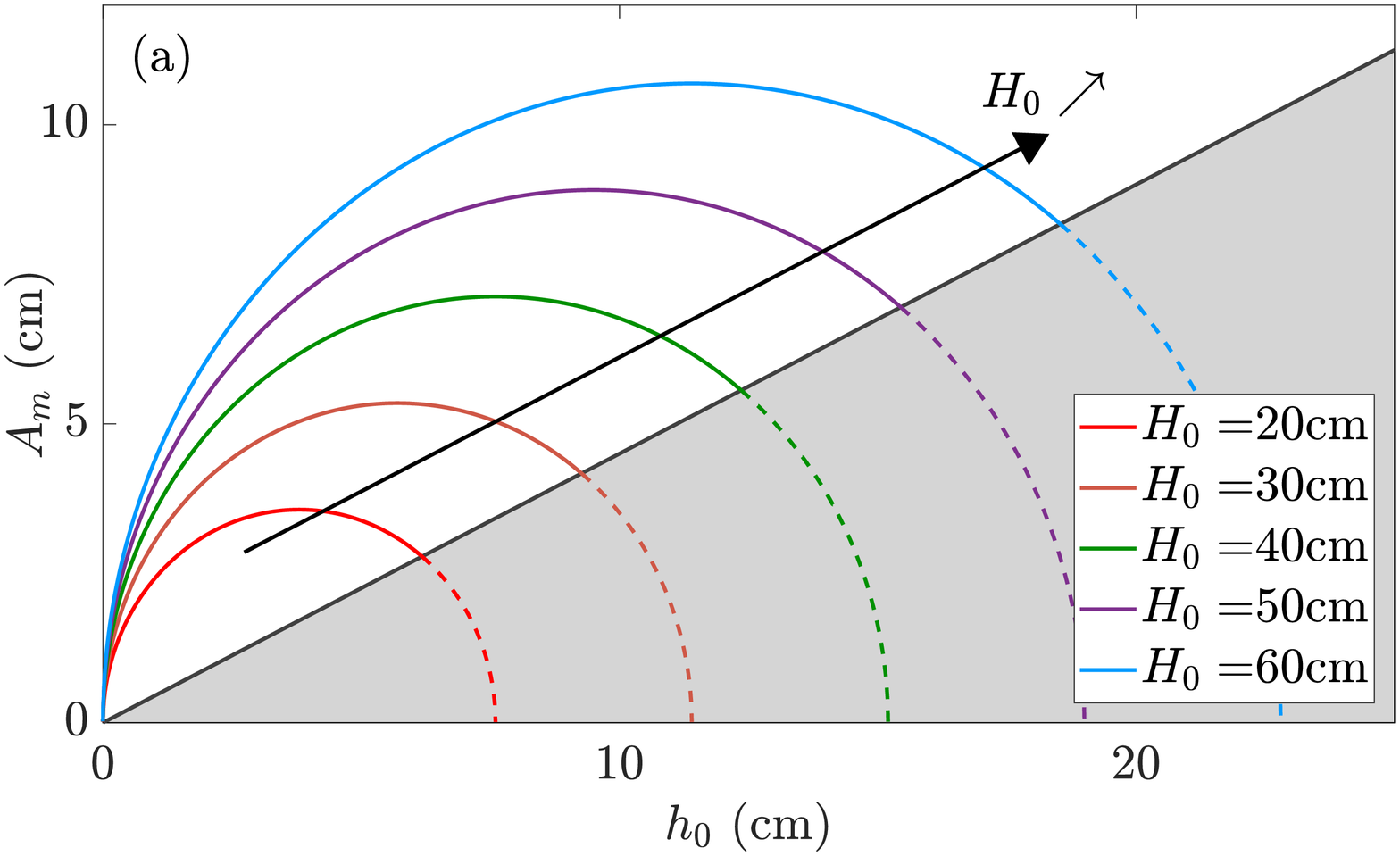}	
	\hfill
	\includegraphics[width=0.49\linewidth]{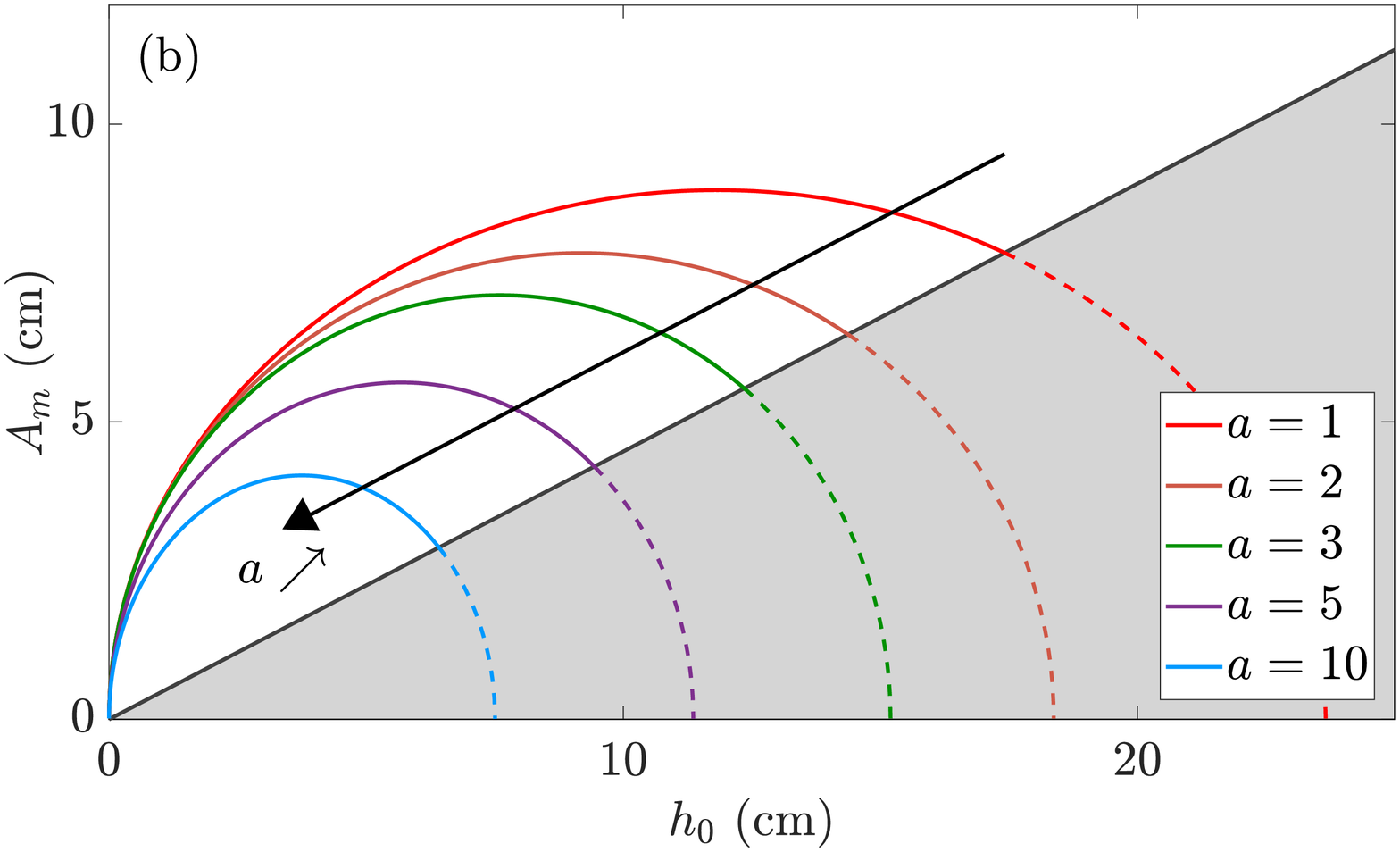}
	\par \vspace{2mm}
	\includegraphics[width=0.49\linewidth]{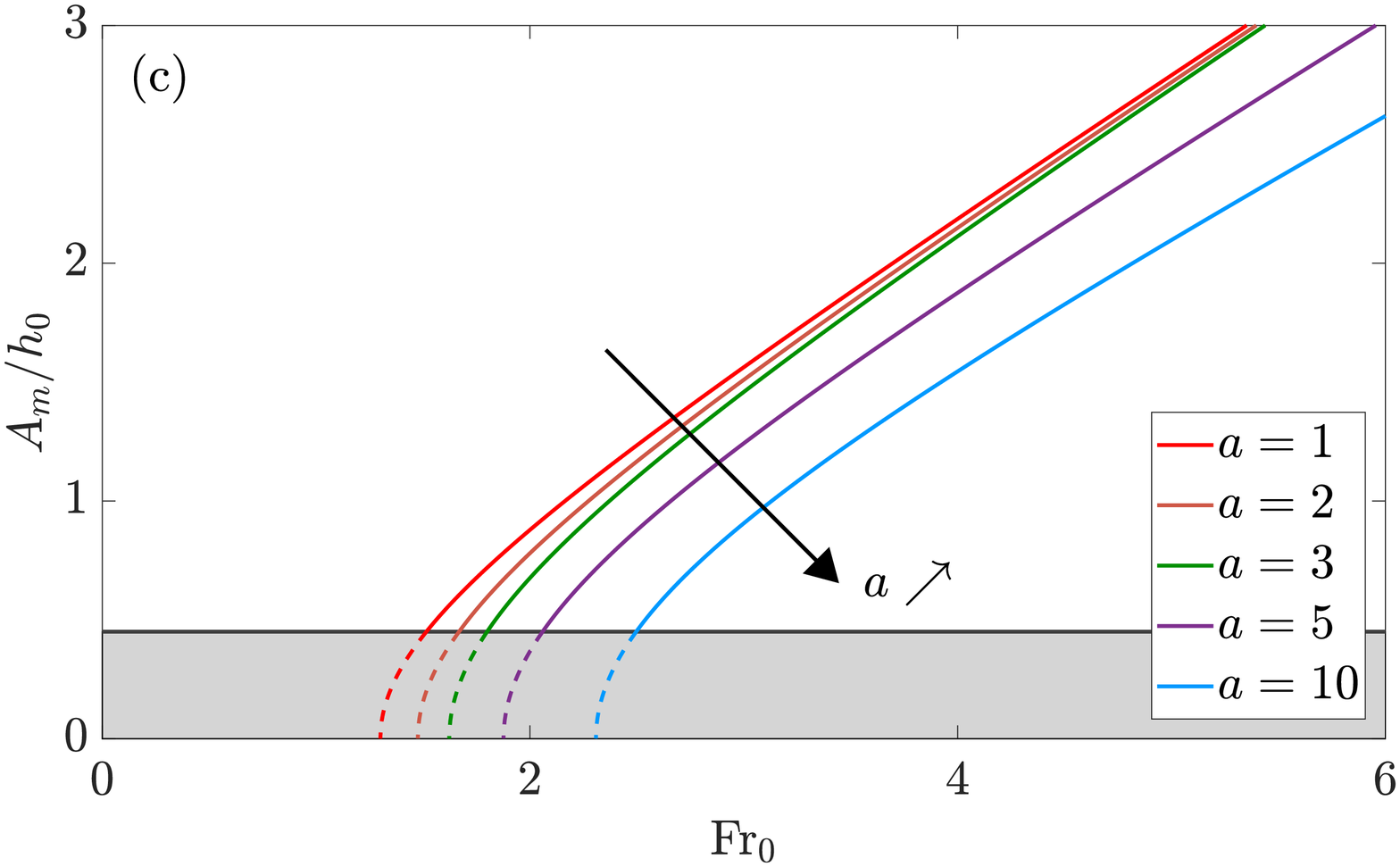}
	\hfill
	\includegraphics[width=0.49\linewidth]{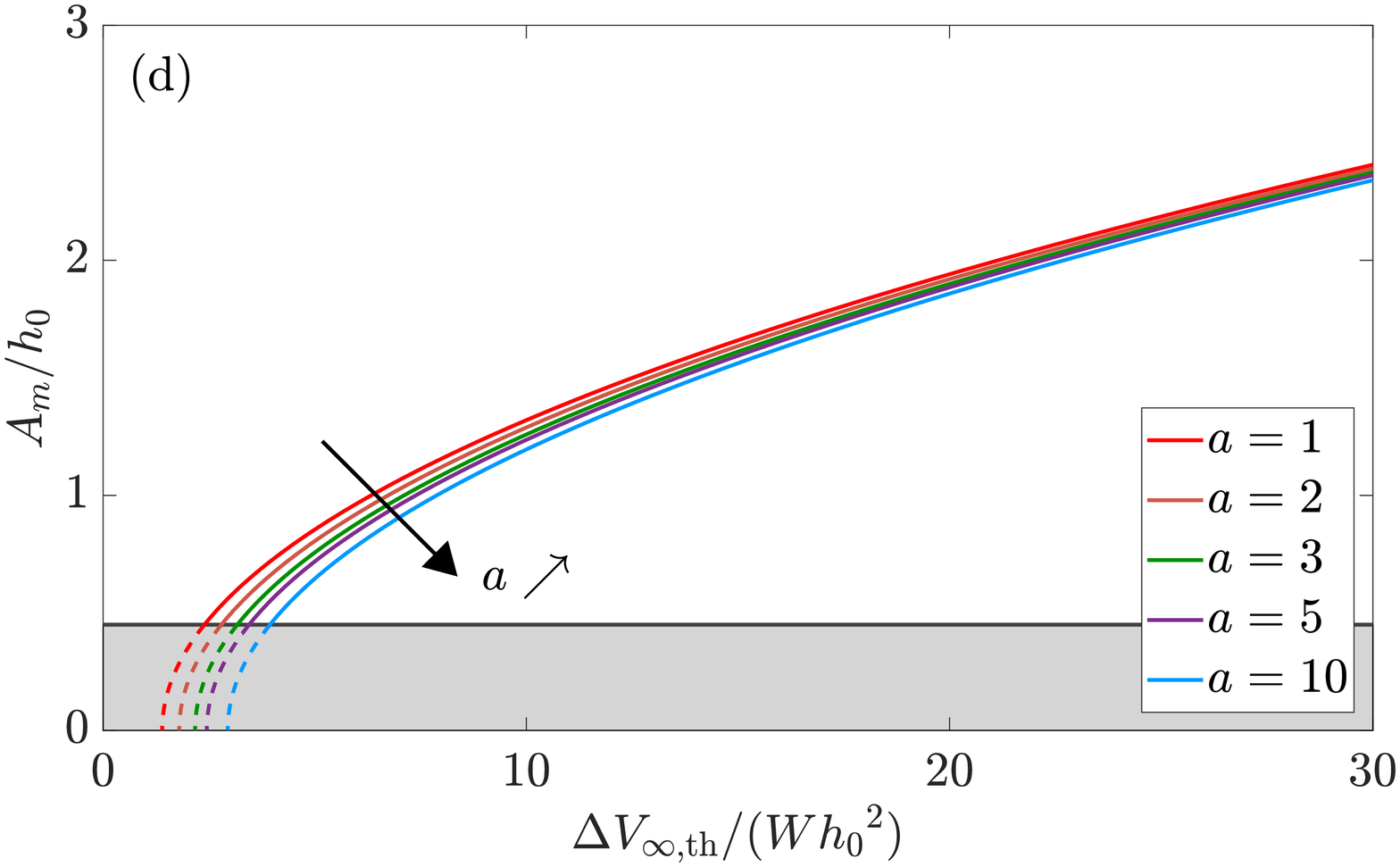}
	\caption{(a)-(b) Maximum wave amplitude $A_m$, given by Eq. \eqref{eq:eq:Am_explicit}, as a function of the water depth $h_0$, for (a) different initial height $H_0$ with the same aspect ratio $a=3$, and (b) different $a$ with $H_0=40$ cm. (c)-(d) $A_m/h_0$ as a function of (c) $\mathrm{Fr}_0$ as given by Eq. \eqref{eq:AmvsFr0}, and (d) $\Delta V_{\mathrm{\infty,th}}/(W{h_0}^2)$ as given by Eq. \eqref{eq:AmvsVimm}, for $H_0=40$ cm and different $a$. Shaded areas are the same as in figure \ref{fig:fig2}.}
	\label{fig:fig4}
\end{figure}

To highlight the influence of the initial parameters on the wave amplitude, figures \ref{fig:fig4}(a)-(b) report the prediction of $A_m$  given by Eq. (5) as a function of $h_0$ for $a=3$ and different values of $H_0$ from 20 to 60 cm [figure \ref{fig:fig4}(a)], and for $H_0=40$ cm and different values of $a$ \PG{from 1 to 10 [figure \ref{fig:fig4}(b)]. In all cases, we observe that $A_m \sim {h_0}^{1/2}$ for the shallowest depths, then reaches a maximum value before decreasing. Note that this maximum, as well as the critical depth at which it occurs, both increase linearly with $H_0$ when $a$ is constant [see Appendix for more details].} The present model reproduces thus well the bell-shaped part of the curves obtained in figure \ref{fig:fig2} which were also observed in previous experimental studies \cite{2017_mulligan,2021a_robbe-saule}.

\begin{figure}[t]
	\centering
	\includegraphics[width=0.49\linewidth]{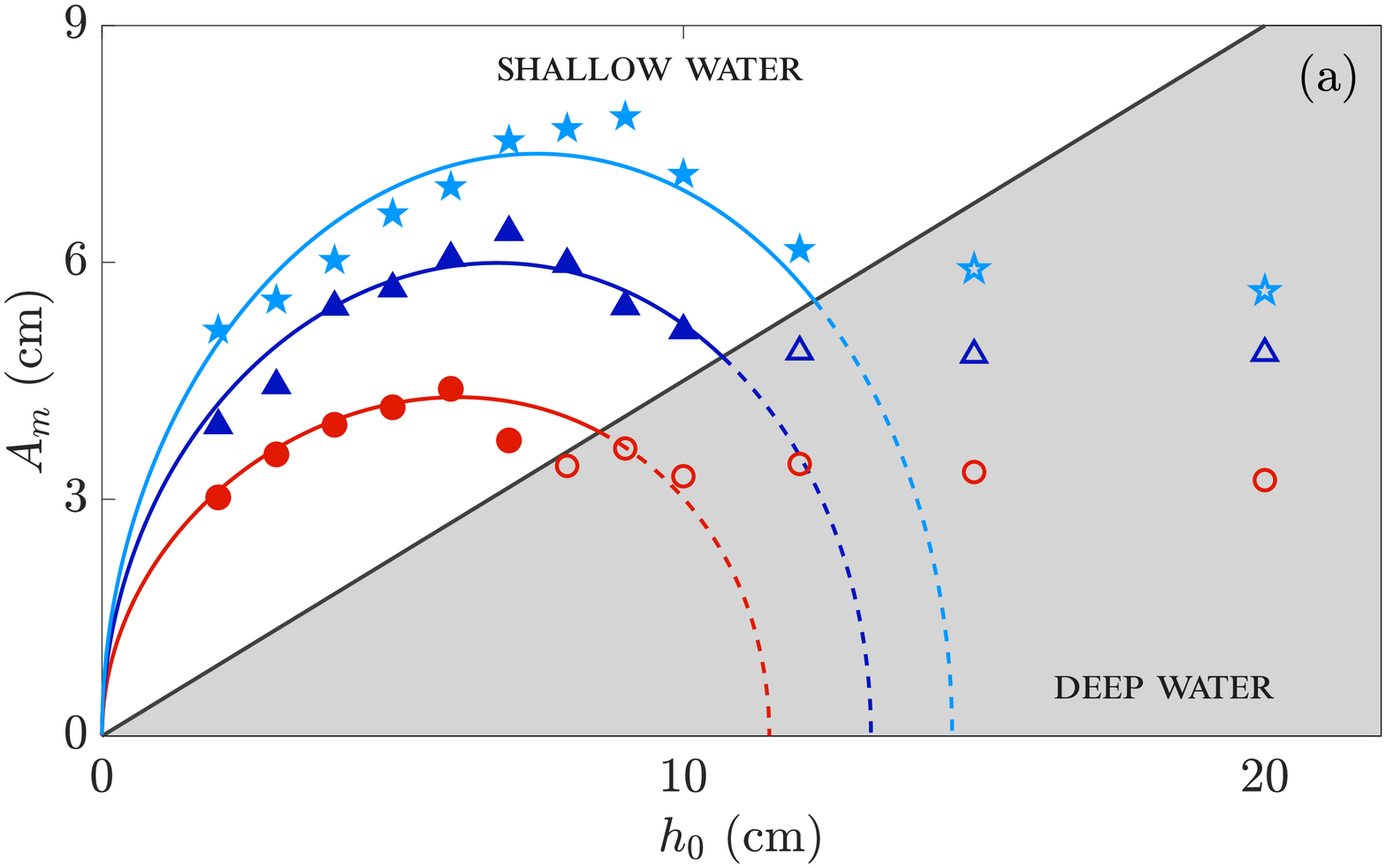}
 	\hfill
	\includegraphics[width=0.49\linewidth]{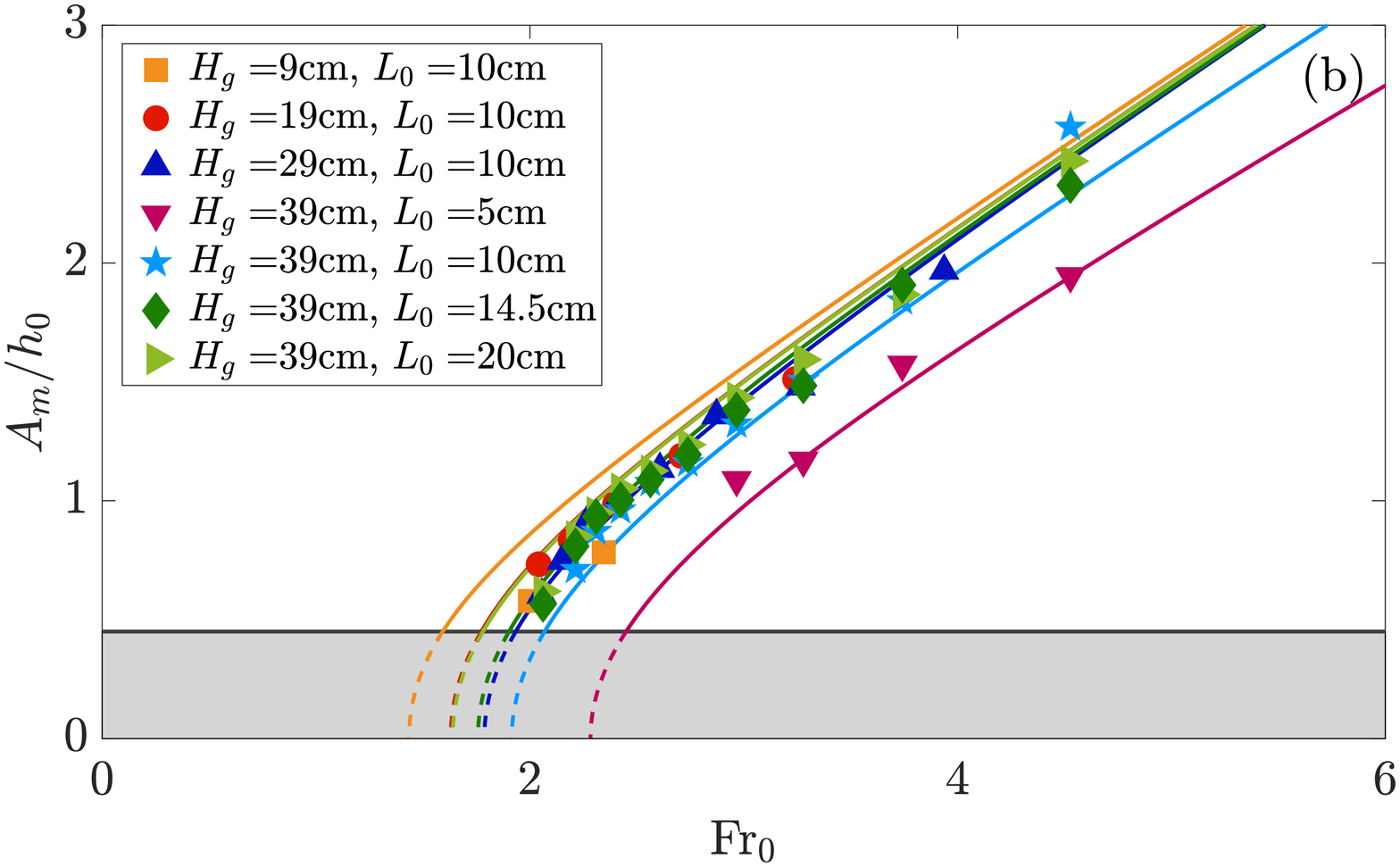}	
	\par \vspace{2mm}
	\includegraphics[width=0.49\linewidth]{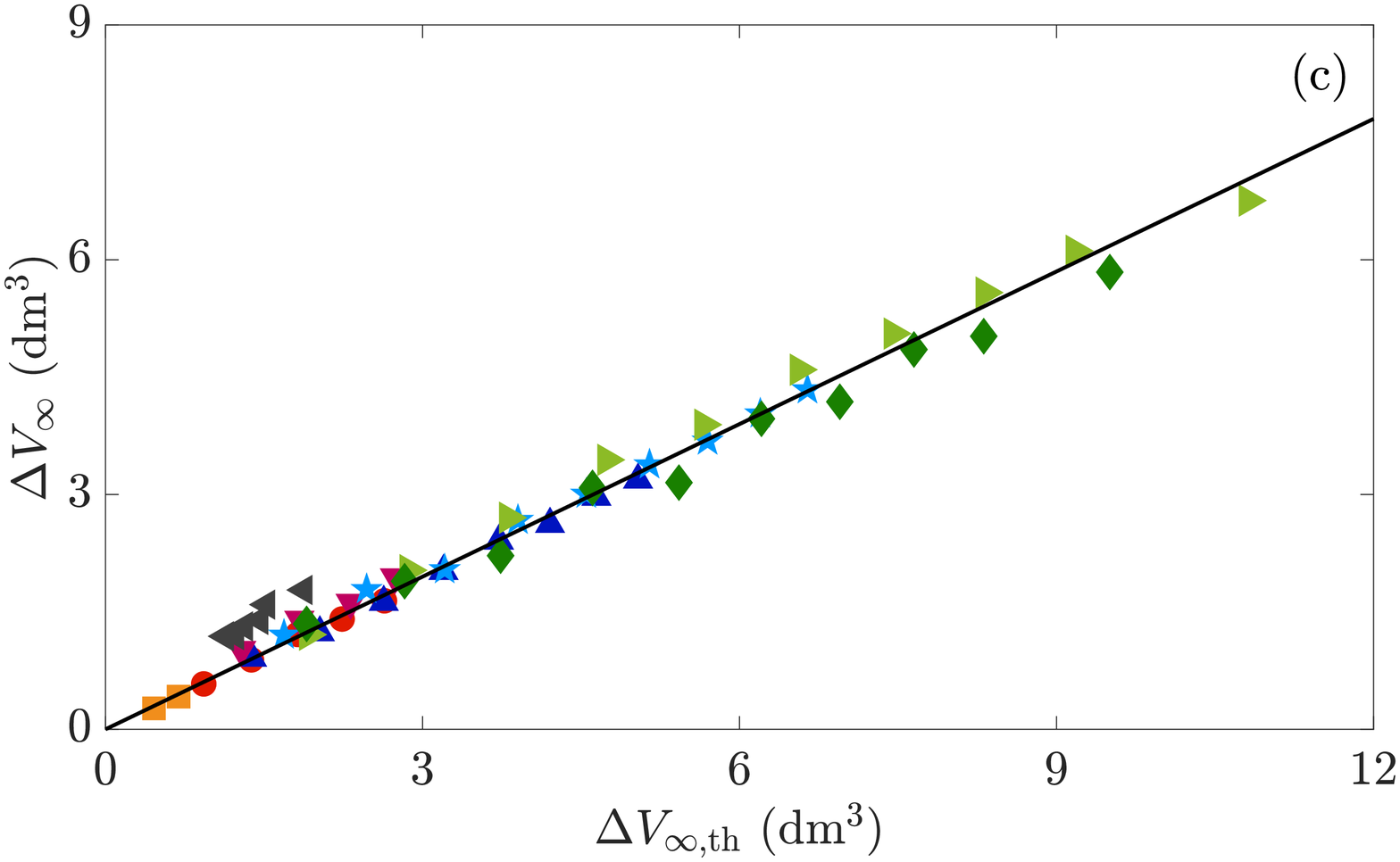}
	\hfill
	\includegraphics[width=0.49\linewidth]{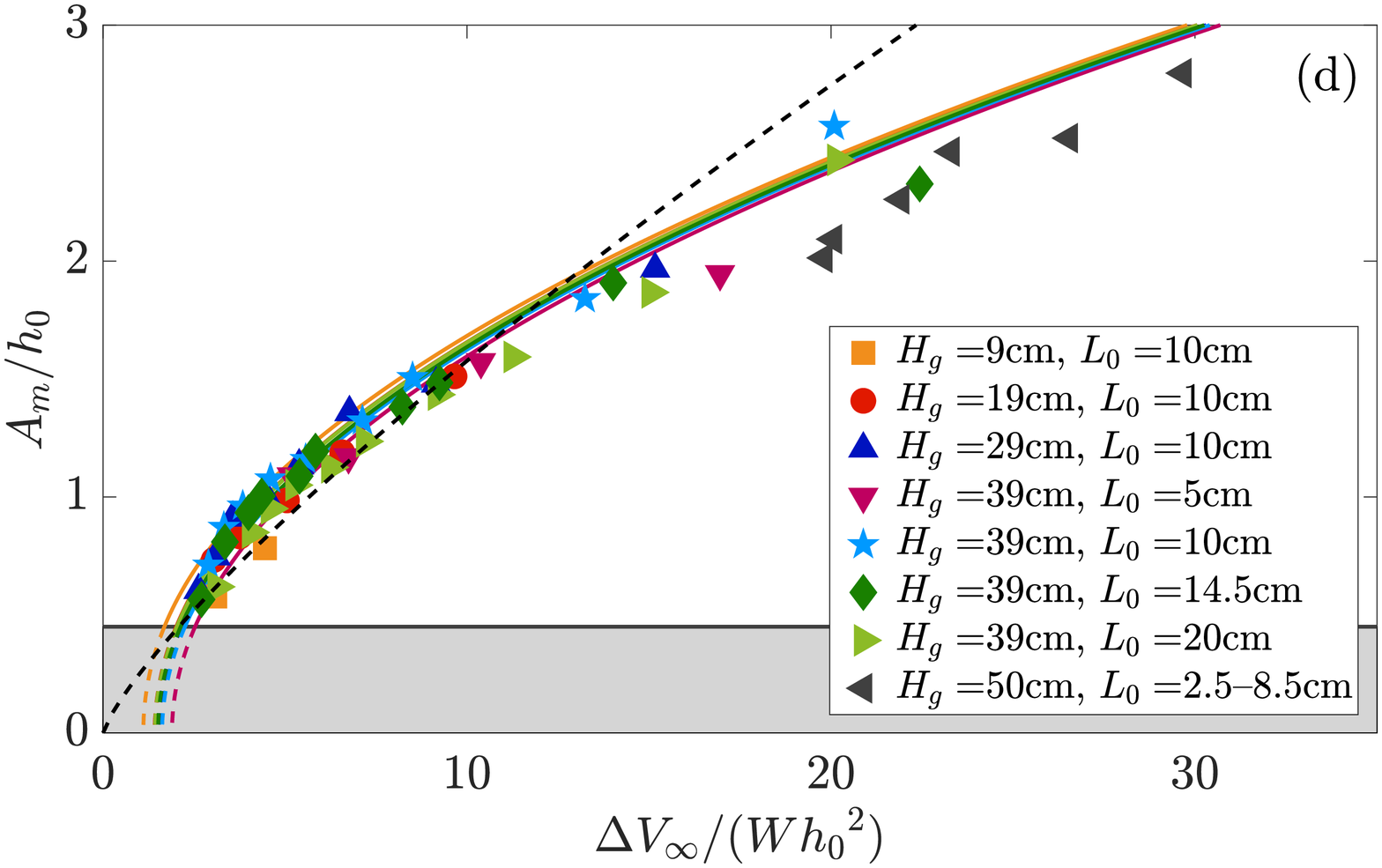}
	\caption{(a) $A_m$ as a function of $h_0$, for the experiments of figure \ref{fig:fig2}. (b) $A_m/h_0$ as a function of the global Froude number $\mathrm{Fr_0}$. (c) Comparison between the experimental and predicted immersed volumes, with (\fullblack) $\Delta V_\infty=0.65\Delta V_{\mathrm{\infty,th}}$. (d) $A_m/h_0$ as a function of the relative volume $\Delta V_\infty/(W{h_0}^2)$ of the immersed deposit, with (\dashedblack) the empirical law $A_m/h_0=0.25 \left[ \Delta V_\infty/(W{h_0}^2) \right]^{0.8}$ from \cite{2021b_robbe-saule}. The colored solid lines in (a), (b), and (d) correspond to the model predictions for the same initial granular columns as in the experiments, and the shaded areas are the same as in figure \ref{fig:fig2}.}
	\label{fig:fig5}
\end{figure}

Let us now explore the prediction of the model for the wave generation as a function of two dimensionless numbers, reflecting either the initial state or the final deposit of the collapse: the global Froude number $\mathrm{Fr}_0=\sqrt{H_0/h_0}$ \cite{2020a_huang,2020_cabrera,2021a_robbe-saule,2022_nguyen} and the relative volume $\Delta V_\infty/(W{h_0}^2)$ of the immersed deposit \cite{2021b_robbe-saule}. Using Eq. \eqref{eq:eq:Am_explicit} straightforwardly leads to 

\begin{equation}
\label{eq:AmvsFr0}
\frac{A_m}{h_0} = 0.45\sqrt{\beta a^{m-1} {\mathrm{Fr}_0}^2 - \frac{1+\beta a^m}{\alpha a^n}}.
\end{equation}

\noindent In addition, within the present model the final immersed deposit would have a trapezoidal shape, of volume $\Delta V_{\mathrm{\infty,th}} = (\Delta L_\infty+\Delta \ell_{\mathrm{\infty,th}}) Wh_0/2$. As a result, Eq. \eqref{eq:Am} leads to

\begin{equation}
\label{eq:AmvsVimm}
\frac{A_m}{h_0} = 0.45\sqrt{ 2\frac{\Delta V_{\mathrm{\infty,th}}}{W{h_0}^2} -\frac{\Delta L_\infty }{h_0} }.
\end{equation}

\noindent The relative wave amplitude predicted by Eqs. \eqref{eq:AmvsFr0} and \eqref{eq:AmvsVimm} is presented  in figures \ref{fig:fig4}(c) and \ref{fig:fig4}(d), respectively, for different initial aspect ratios \PG{from 1 to 10} and $H_0=40$ cm. In both cases, $A_m/h_0$ increases monotonically \PG{with a sublinear shape. Note that the influence of the aspect ratio on the relative wave amplitude is almost imperceptible when varying the rescaled volume of the immersed deposit in figure \ref{fig:fig4}(d), but is non-negligible when the global Froude number is varied, as in figure \ref{fig:fig4}(c).}

\medskip

\noindent \textbf{{Comparison with the experiments.}} In figure \ref{fig:fig5}(a), the prediction of the wave amplitude $A_m$ using Eq. \eqref{eq:eq:Am_explicit} is compared to the experiments from figure \ref{fig:fig2}. The numerical prefactors found by the best fit for each initial geometry are in the range [0.42--0.49], \textit{i.e.}, very close to the value 0.45 of Eq. \eqref{eq:eq:Am_explicit}. A good agreement is observed in shallow water conditions, as the bell-shaped curves are correctly reproduced. \PG{As expected, the present model fails in predicting the plateau values observed for deep water waves ($A_m \lesssim 0.45~h_0$) since the model is not valid under these conditions.} The relative wave amplitude $A_m/h_0$ is also reported as a function of $\mathrm{Fr}_0$ in figure \ref{fig:fig5}(b), alongside the model curves obtained for identical initial granular columns.  \PG{All experimental data are within the collection of model curves from Eq. \eqref{eq:AmvsFr0}, with a sublinear increase of $A_m/h_0$ with $\mathrm{Fr}_0$ for a given $a$. The influence of the aspect ratio is also well captured in the investigated range, as shown by the experiments (\textcolor{color_4}{$\blacktriangledown$}) with the most slender initial columns, that are significantly below the other data (corresponding to lower values of $a$).} Therefore, at first order, the global Froude number $\mathrm{Fr}_0$ governs the relative wave amplitude, but with a noticeable effect of the initial aspect ratio. \PG{It is also interesting to consider the correlation between the relative wave amplitude and the relative volume of immersed deposit. To do so, we first compare the model prediction for the volume of the immersed deposit $\Delta V_{\mathrm{\infty,th}}$ with the experimental measurements of $\Delta V_\infty$. Figure \ref{fig:fig5}(c) shows that there is a good correlation between them.} A linear trend of slope 0.65 fits well the data, showing that the predicted volume overestimates the measurements in a systematic way\PG{, here again because the final immersed deposit is not perfectly trapezoidal}. Using this result, it is possible to compare $A_m/h_0$ to $\Delta V_\infty/(W{h_0}^2)$ for all experiments and to draw the corresponding model curves, as shown in figure \ref{fig:fig5}(d). The agreement between the experiments and the predictions allows one to explain the strong link existing between $\Delta V_\infty/(W{h_0}^2)$ and the generated wave, which was previously reported \cite{2021b_robbe-saule}. \PG{This correlation is a striking result, that may appear surprising, but one needs to keep in mind that the final deposit is the result of the gravity-driven collapse and is thus reminiscent of its dynamics.} The empirical equation for the wave amplitude given in \cite{2021b_robbe-saule} is also reported in figure \ref{fig:fig5}(d), to highlight that the present model better captures the \PG{experimental} data in \PG{the whole range of} shallow water conditions.



\medskip
\noindent \textbf{{Conclusion.}} The amplitude $A_m$ of the impulse wave generated by the collapse of a granular column in a shallow water of depth $h_0$ can be predicted by combining the spreading dynamics of the grains \cite{2021c_sarlin}, relating the initial parameters of the column to the local Froude number $\mathrm{Fr}_f$ \PG{based on the advancing granular front at the water surface}, and the wave hydrodynamics linking $\mathrm{Fr}_f$ to $A_m$ \cite{2021b_sarlin}. In this situation, the spreading motion of the grains indeed behaves as a peculiar piston, whose velocity is coupled with its stroke. The present model explicits the evolution of $A_m$ as a function of the initial parameters $H_0$, $L_0$, and $h_0$. In addition, it highlights the important role played by the global Froude number $\mathrm{Fr}_0=\sqrt{H_0/h_0}$ and the relative volume $\Delta V_\infty/(W{h_0}^2)$ of the immersed deposit. It explains why these different dimensionless numbers have been observed to play a key role in previous studies \cite{2020a_huang,2021a_robbe-saule,2021b_robbe-saule,2022_nguyen}. 

\PG{It is worth noting that the numerical prefactors obtained in the model slightly depend on the considered granular matter. Therefore, one would need to  adjust their values to apply the present model to other materials. Within these minor adjustments, this predictive model should work in similar configurations, for instance for the ones explored experimentally \cite{2020a_huang,2020_cabrera} or numerically \cite{2022_nguyen} in previous studies, provided that the shallow water condition is fulfilled. For deep water conditions, a predictive model remains to be developed. The present model should be adapted accordingly for the topographies and bathymetries that are specific in the field, by expressing the maximum velocity of the granular front in the impact region, to derive the relevant Froude number in that case. For post-mortem analysis in the field based on the measurements of the final immersed deposits, the present model could be used without significant adjustments. Indeed, there is a trace of the gravity-driven dynamics in the final deposit, as shown by the good correlation of Eq. (7) with all the data for very different columns with a large range of investigated aspect ratios. In order to investigate 3D effects that may be important in real cases, in particular in the collapse of volcanic islands such as the 2018 Krakatau event \cite{2019_grilli}, the study of waves generated by the collapse of cylindrical columns instead of rectangular ones would be of great interest. Indeed, the scaling laws for the final deposit are known to be different in this case \cite{2004_lajeunesse,2004_lube}. In addition, a 3D context would make the wave hydrodynamics more complex.}



\PG{\section*{Appendix : Critical depth for a maximum wave amplitude}}
\label{SecVI}

\PG{It is interesting here to provide a discussion on the critical water depth at which the maximum wave amplitude is reached, for the collapse of a granular column of a given aspect ratio, as observed in the bell-shaped curves of Figs. 5(a)-(b). From Eqs (3)-(4), one can  obtain that the critical depth is given by $h_0^c=\Delta L_\infty H_\infty /(2L_\infty)$, which leads to the corresponding amplitude $A_m^c \simeq 0.45 \sqrt{\Delta {L_\infty}^2 H_\infty/(4L_\infty)}$. For a column with a given aspect ratio $a$, leading to a given deposit, we can infer from these expressions that both $h_0^c$ and $A_m^c$ increase linearly with the initial height $H_0$ of the column, as can be observed in Fig. 5(a).}  To illustrate this point, fom Eqs. (\ref{eq:final_morphology}) we can write $h_0^c/H_0$ in terms of the initial aspect ratio $a$ of the granular column in the following manner:

\begin{equation}
\label{eq:h_0cr}
\begin{aligned}
	\frac{h_0^c}{H_0} \simeq \left\{ \begin{array}{ll}
	\displaystyle \frac{0.83a}{1+1.65a} \qquad \qquad & \mathrm{for\ }a \lesssim 0.93,\\[8pt]
	\displaystyle \frac{0.78a^{1/3}}{1+1.65a}\qquad \qquad & \mathrm{for\ }0.93 \lesssim a \lesssim 3,\\[8pt]
	\displaystyle \frac{1.13}{1+2.38a^{2/3}} \qquad \qquad & \mathrm{for\ }a \gtrsim 3.
	\end{array}\right.
\end{aligned}
\end{equation}


\noindent \PG{These expressions thus show that $h_0^c$ is proportional to $H_0$ when the aspect ratio of the column is kept constant.}



\begin{figure}[t]
	\centering
	\includegraphics[width=0.49\linewidth]{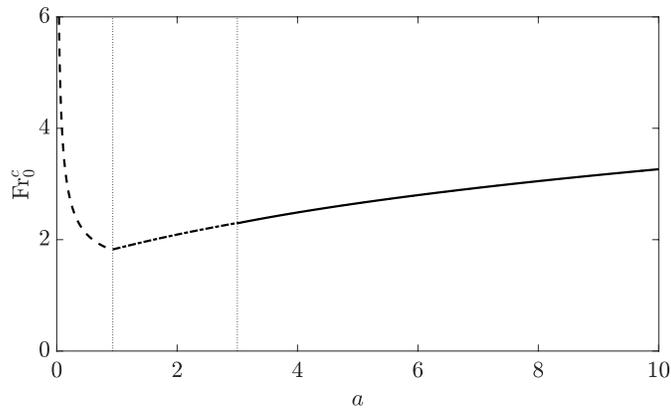}
	\caption{Critical Froude number $\mathrm{Fr}_0^c=\sqrt{H_0/h_0^c}$ at which the maximum wave amplitude is reached for the collapse of granular column, as a function of the initial aspect ratio $a$. \PG{The line corresponds to Eqs. \eqref{eq:h_0cr} for the three domains separated by the two vertical dotted lines at $a=0.93$ and $a=3$}.}
	\label{fig:fig7}
\end{figure}

\noindent The associated critical Froude number $\mathrm{Fr}_0^c=\sqrt{H_0/h_0^c}$ is presented in figure \ref{fig:fig7} as a function of the aspect ratio of the column. It shows a nontrivial evolution with $a$, with a sharp decrease for low aspect ratio ($a \lesssim 0.93$), and an increasing region for higher values of $a$. It means that, for aspect ratios below 0.93, the maximum value for the amplitude of the wave is reached for $h_0^c$ small when compared to the initial column height $H_0$. Around $a \simeq 0.93$, the curve reaches a global minimum where $\mathrm{Fr}_0^c \simeq 1.8$, so that $h_0^c \simeq H_0/3.24$: It is the highest possible value of $h_0^c$ when varying the initial aspect ratio of the column. Then, increasing $a$ beyond this point again implies that the critical wave amplitude occurs with $h_0^c$ increasingly small when compared to $H_0$.\\

\begin{acknowledgments}
The authors are grateful to J.~Amarni, A.~Aubertin, L.~Auffray and R.~Pidoux for the elaboration of the experimental setup, and report no conflict of interest.
\end{acknowledgments}

\bibliography{bibliography}

\end{document}